\documentclass[12pt]{article}
\usepackage{graphicx,graphics,cite}
\begin{document}
\title{\bf Machian Origin of the Entropic Gravity and Cosmic Acceleration}
\author{{\bf Merab Gogberashvili}\\
Andronikashvili Institute of Physics\\
6 Tamarashvili str., Tbilisi 0177, Georgia\\
and\\
Javakhishvili State University\\
3 Chavchavadze av., Tbilisi 0128, Georgia\\
{\sl E-mail: gogber@gmail.com}
\\
\\
{\bf Igor Kanatchikov}\\
National Quantum Information Centre of Gda\'nsk, \\
ul. W\l. Andersa 27, 81-824 Sopot, Poland \\
{\sl E-mail: kai@fuw.edu.pl}
\date{December, 2010}
}
\maketitle
\begin{abstract}
We discuss the emergence of relativistic effects in the Machian universe with a global preferred frame and use thermodynamic considerations to clarify the origin of gravity as an entropic force and the origin of dark energy/cosmic acceleration as related to the Hawking-Unruh temperature at the universe's horizon.


\end{abstract}

\vskip 0.5cm


\section{Introduction}

There exist three types of field theoretical descriptions successfully working in their own areas: quantum mechanics, quantum field theory, and general relativity. However, problems arise, when one tries to establish bridges between these descriptions. There are well known difficulties in the foundations of relativistic quantum mechanics, the definition of quantum fields in curved space-times, or quantization of gravity. A new hope to establish a unifying approach to main field theoretical descriptions is provided by the approach appealing to the notions of thermodynamics. For instance, thermodynamic arguments have been used in the black hole physics \cite{BH}, discovery of Unruh temperature \cite{Unr}, establishment of the AdS/CFT correspondence \cite{AdS}, derivation of the Einstein \cite{Ein} and Maxwell \cite{Wang} equations, in the recent attempt to interpret the Newtonian gravity as an entropic force \cite{Ver}, and other discussions of the 'emergent gravity' \cite{Pad,Hu}. Besides, the analogies with classical statistical mechanics and thermodynamics have been underlying some recent discussions of the foundations of quantum mechanics \cite{Wet,Gro}.

It seems that the relations between gravity and thermodynamics may open a possibility to overcome the intrinsic limitations of purely geometric treatments of gravity. Those limitations are obvious in the nature of the Einstein equations itself, which make the borderline between the matter and geometry too vague, if not conventionalist. As a consequence, numerous extreme attempts to view the matter as geometry or treat the space-time as a kind of material medium can be found in the literature.

The main difficulty of General Relativity is related to the description of gravitational energy. This problem has been considered in numerous discussions of the energy-momentum pseudo-tensors (for the recent review, see \cite{Sza}). Being based on the principle of equivalence, General Relativity requires that in a properly chosen coordinate system the Christoffel symbols can vanish in a vicinity of any given point. It essentially implies a possibility to nullify the gravitational energy of a body at any given point, and hence a violation of the energy conservation law. However, if we insist that the energy conservation law is a universal law which must be valid also for the gravitational energy, then this feature seems to be not satisfactory.

The situation reminds us a problem from the history of classical thermodynamics: the apparent violations of the energy conservation law in the heat transfer processes. In order to save the energy conservation law the concept of entropy was introduced which characterizes the statistics of microstates and is responsible for the dissipation of heat in the system. The resulting first law of thermodynamics asserts that the entropy change compensates the energy loss of a hot body. Correspondingly, according to the second law, any process of heat transfer is always accompanied by the increase of entropy. It is essential that the thermodynamic description, being deduced merely from the fact that the object can be heated, is insensitive to the details of the dynamics of microstates.

Besides, the observations of cosmic acceleration indicate that our understanding of classical gravity may be fundamentally incomplete, because the description of this phenomenon within General Relativity requires an {\it ad hoc} matter with exotic properties. This is also reminiscent of an early history of the theories of heat which were using unphysical substances such as phlogiston or caloric until it was remedied by the concepts of thermodynamics and the mechanical theory of heat.

There are evidences that it may be more appropriate to understand gravity as a local change of temperature, instead of describing it using a purely geometrical language \cite{Pad,Hu}. The advantage of this approach would be the exact conservation of energy in the spirit of the first law of thermodynamics. In this way the apparent disappearance of gravitational energy in geometric formulations of gravity could be avoided.

So far the thermodynamic relations in the context of field theories have been appearing mostly as heuristic analogies, because the origin of microscopic degrees of freedom, which would give rise to the thermodynamic quantities, is unclear. Besides, the thermodynamic approaches face a problem of explaining the origin of the relativity principle, because the state of the thermodynamic equilibrium in the universe implies the appearance of a global preferred reference frame, thus contradicting even the Galilean principle of relativity.

In order to assert that gravity can be really obtained from thermodynamics, the above-mentioned issues should be clarified. In this paper we address those problems using the Machian model of the universe introduced earlier in \cite{Gog1,Gog2,Gog3,Gog4}.

We proceed as follows. In Sect. 2 we recall basic ingredients of the Machian model of the universe and in Sect. 3 show how the elements of Special and General relativity theory can be derived from what we call the Machian energy balance conditions. Then, in Sect. 4, the notions of temperature and entropy are introduced in the Machian model and the origin of gravity as an entropic force is clarified. The formula for the temperature on the universe's horizon is obtained in Sect. 5, and the corresponding acceleration of the horizon is found to be consistent with the observed cosmic acceleration and dark energy. The conclusions are found in Sect. 6.


\section{The Machian Model of the Universe}

Let us recall that the Machian model introduced in our previous papers \cite{Gog1,Gog2,Gog3,Gog4} postulates a non-local collective gravitational interaction of all particles inside the universe's horizon, as a consequence of which all $N$ particles in the universe are "gravitationally entangled" and form a unified statistical ensemble. The universal constant of the speed of light, $c$, originates in the non-local potential of the whole universe, $\Phi$, acting on any particle of the world ensemble:
\begin{equation} \label{Phi}
c^2 = - \Phi = \frac{2M_U G}{R}~,
\end{equation}
where $M_U$ and $R$ are the total mass and the radius of the universe, respectively, and $G$ is the Newton constant. Note that it is due to the cosmological principle, i.e. the isotropy and homogeneity of the universe at the horizon scale $R$, that $c$ has the status of universal constant. Equation (\ref{Phi}) allows us to formulate the Mach principle which relates the origin of inertia of a particle, or its rest energy, to its interactions with the whole universe:
\begin{equation} \label{mPhi}
E = -m\Phi = mc^2~,
\end{equation}
where $m$ is the mass parameter describing the particle's inertia, which in general is not constant.

Several interpretations of Mach's principle can be found in the literature \cite{Mach}. The usual interpretation, which leads to the anisotropy of the rest mass of particles due to the influence of nearby massive objects (like the Galaxy), has been ruled out by experiments \cite{exp}. In fact, the naive Machian conjecture that the mass parameter can be altered by "distant stars" is based on the assumption that kinematics, or local space-time, is independent from the surrounding matter. However, the lesson of General Relativity has been exactly that the description of space-time geometry, and hence the kinematics itself, depends on the distribution of energy. Therefore, the influence of the whole universe on the local physics, or Mach's principle in its utmost generality, should not be described in terms of the mass parameters alone, but rather in terms of more fundamental quantities such as action or energy.

Our Machian model is based on the energy balance conditions such as (\ref{mPhi}) below which provide a better framework for the treatment of the influence of the universe on the local physics and the description of dynamical and kinematical parameters. Note that (\ref{mPhi}) is the simplest Machian energy balance relation which is written here for a particle at rest with respect to the preferred reference frame of the universe, since it takes into account only the Machian interaction of the particle. More complicated energy balance equations will be discussed in what follows.

Equation (\ref{mPhi}) takes into account the contribution of the collective gravitational interactions between all $N$ particles inside the horizon. Namely, since each particle interacts with all other $(N-1)$ particles, and the mean separation in the interacting pairs is $R$, the total Machian energy consists of $N(N-1)/2$ terms of magnitude $\approx 2G m^2/R$. Then, for very large $N$, the Machian energy of a single particle which interacts with the total Machian potential of the universe $\Phi$ is given by:
\begin{equation} \label{E}
E \approx N^2 \frac {Gm^2}{R} ~.
\end{equation}
Correspondingly, the contribution of the collective Machian interactions to the total mass of the universe is:
\begin{equation} \label{M=N2m}
M_{Mach} \approx \frac{1}{2} N^2 m ~,
\end{equation}
so that the total mass of the universe $M_U\sim M_{Mach}$ is of the order of $N^2m$ and not $\approx Nm$.

Let us emphasize that (\ref{Phi}) is equivalent to the critical density condition in relativistic cosmology:
\begin{equation} \label{rho}
\rho_{c} = \frac{3M_U}{4\pi R^3} = \frac{3H^2}{8\pi G} ~,
\end{equation}
where $H \sim c/R$ is the Hubble constant. Using the observed values of $c$, $G$ and $H$ in (\ref{rho}) or in (\ref{Phi}), we can estimate the total mass of the universe:
\begin{equation} \label{M}
M_U \sim \frac{c^3}{2 GH} \approx 10^{53}~ kg~.
\end{equation}

Because of the finite number of particles inside the horizon and the existence of the maximal speed $c$ any movement of particles of the "gravitationally entangled" world ensemble results in a delayed response time of the whole ensemble $\Delta t$. Any mechanical process in the world ensemble will be accompanied by the exchange of at least the minimal amount of action $A = m c^2 \Delta t$, which we identify with the Planck's action quantum \cite{Gog3}:
\begin{equation} \label{A}
m c^2 \Delta t =: \hbar ~.
\end{equation}
The response time $\Delta t$ of the ensemble to the particle's motion is estimated as follows:
\begin{equation} \label{Deltat}
\Delta t \sim \frac{R}{Nc} \sim \frac{1}{NH}~.
\end{equation}
Here we take into account that due to the non-locality of the world ensemble its members can interact independently from their separation so that the effective mean distance between particles in the model is $\sim R/N$, i.e. much shorter than e.g. the mean separation $\sim R / N^{1/3}$ in a dilute gas.

From (\ref{M}), (\ref{Deltat}) and (\ref{A}) we can estimate the total action of the universe:
\begin{equation} \label{A_U}
A_U = \frac{M_U c^2}{H} \approx \frac{N^3}{2} A \approx\frac{N^3}{2} \hbar ~,
\end{equation}
and the number of typical particles in world ensemble:
\begin{equation} \label{N}
N \approx \left(\frac{2 A_U}{A} \right)^{1/3}
\approx \left(\frac{2M_U c^2}{\hbar H} \right)^{1/3}
\approx 10^{40} ~.
\end{equation}
Using this value and (\ref{M=N2m}) and (\ref{M}), we can also estimate the mass of a typical particle of the ensemble:
\begin{equation} \label{m}
m \approx {\frac{2 M_U}{N^2}} \approx 2 \times 10^{-27}~ kg
\approx 1 ~ GeV~c^{-2}~,
\end{equation}
which appears to be of the order of magnitude of the proton mass. This estimation is also consistent with (\ref{Deltat}), as
\begin{equation} \label{Deltat=}
\Delta t \approx \frac{1}{NH}
\sim 10^{-23}~s \sim  \hbar ~ GeV^{-1}~.
\end{equation}
Hence, a typical stable heavy particle, the proton, can be considered as a typical particle forming the gravitating world ensemble in our simplified one-component Machian universe.

The energy balance equations like (\ref{mPhi}) exhibit the exact conservation of energy in the Machian universe, and thus they may help to avoid the problems with energy in Einstein's General Relativity. Those energy balance conditions assume that the non-local Machian gravitational interaction with the universe is the source of all kinds of local energy of a particle, i.e. the total (Machian plus local) energy of any object vanishes, if the gravitational energy is considered as negative and all other forms of energy are assumed to be positive. For example, (\ref{mPhi}) is equivalent to:
\begin{equation} \label{balance}
mc^2 + m\Phi = 0~.
\end{equation}
Hence, the total energy of the whole universe also vanishes. This point of view appears to be preferable in cosmology \cite{Haw}, since in this case the universe can emerge without violating the energy conservation.

Note that the energy balance condition (\ref{balance}) actually is equivalent to the equation of state for the dark energy in standard cosmology:
\begin{equation}
\rho + p = 0~,
\end{equation}
where $\rho$ is the energy density. In our case, the exotic minus sign for the pressure $p$ has a natural explanation: it appears since the Machian gravitational potential of the universe $\Phi$ is negative. Below in Sect. $5$, using thermodynamic arguments, we show that the energy balance condition (\ref{balance}), when written for the whole universe, leads to the observable value of the cosmic acceleration.


\section{Relativity and the Preferred Frame}

The energy balance conditions such as (\ref{balance}) provide a framework for the treatment of the Machian influence of the universe on the local physics and the description of dynamical and kinematical parameters. For example, in \cite{Gog2} it was demonstrated that, in spite of the existence of a preferred frame, the approach based on the Machian energy balance equations is able to imitate basic features of the Special Relativity theory. Namely, the relativity principle emerges from the fact that in the homogeneous Machian universe there exists a class of privileged observers which have constant velocities with respect to the preferred frame, for which the universe looks spherically symmetric (see the standard definition of inertial frames e.g. in \cite{Wei}). However, in order to introduce the relativity principle into the model with a preferred frame, we have to pay with the velocity dependence of the inertia of particles. To show this, let us define the total energy of a particle in an inertial frame by adding into the energy balance equation, (\ref{mPhi}) (which was written for the particle at rest with respect to the preferred frame), the term corresponding to the particle's kinetic energy $E_{kin}$:
\begin{equation} \label{Etot}
E_{tot} = E + E_{kin} ~.
\end{equation}
Then we can define, similarly to (\ref{mPhi}), the mass of the moving particle as follows:
\begin{equation} \label{E/Phi}
m = - \frac{E_{tot}}{\Phi} ~,
\end{equation}
which is obviously velocity dependent. Hence, the Machian influence of the universe in this case manifests itself in the velocity dependence of the inertia.

On the other hand, the number of particles of the world ensemble inside the horizon of an inertial observer, and thus the Machian energy $E$, is the same for all inertial observers, i.e.
\begin{equation}
E = E_{tot} - E_{kin} = const~,
\end{equation}
or
\begin{equation} \label{dE}
dE = c^2 dm - d E_{kin} = 0~.
\end{equation}
Then, using the Hamilton's equations and the definition of momentum $p^i = mv^i$, the latter equation can be transformed as follows:
\begin{eqnarray} \label{cal}
c^2 dm - d E_{kin} &=& c^2 dm - v^i dp_i = \left( c^2 - v^2 \right) dm - \frac m2 dv^2 =\\
&=& m \left( c^2 - v^2 \right) d \left( \ln m\sqrt{1 - \frac{v^2}{c^2}}\right) = 0~. \nonumber
\end{eqnarray}
Consequently, the quantity
\begin{equation} \label{m0}
m_0 := \frac{m}{\gamma}~,
\end{equation}
where
\begin{equation} \label{gamma}
\gamma = \frac{1}{\sqrt{1- v^2/c^2}}
\end{equation}
is the standard Lorentz factor, is constant, and hence it can be interpreted as a mass parameter of a particle which is valid in any inertial frame (also known as the rest mass). Thus, in the Machian universe the well-known relativistic effect of the increase of inertia with velocity has a dynamical nature and appears as a consequence of the Machian energy balance condition, (\ref{Etot}), rather than the Lorentz transformations.

The Lorentz transformations and the Minkowskian space-time picture can be also derived as consequences of the Machian consideration where the physical fact of the constancy of the Machian energy, (\ref{dE}), replaces the formal geometric requirement of the invariance of 4-intervals. Correspondingly, within the Machian model of the universe all other well-known relativistic effects, such as the length contraction and time dilation, also have a dynamical nature.

For example, in the formula of the linear momentum:
\begin{equation} \label{p}
p^i = m v^i = m_0\gamma v^i~,
\end{equation}
the factor $\gamma$ can be attributed to $dt$ in $v^i = dx^i/dt$ rather than to the mass $m$, and then the notion of the "proper time" $\tau$ can be introduced:
\begin{equation} \label{tau=t}
d\tau = \frac{dt}{\gamma}~.
\end{equation}

In terms of the rest mass (\ref{m0}) the expression for the total energy (\ref{Etot}) takes the form:
\begin{equation} \label{E-relat}
E_{tot} = mc^2 = m_0\gamma c^2 = m_0c^2 + T_{kin} ~,
\end{equation}
where
\begin{equation}
T_{kin} = \frac{m_0v^2\gamma^2}{1 + \gamma} = \frac{p^2}{m_0(1 + \gamma )} ~,
\end{equation}
is the relativistic kinetic energy. Now, the identity
\begin{equation}
\gamma^2 c^2(c^2 - v^2) = c^4~
\end{equation}
multiplied by $m_0$ yields the standard 4-dimensional relation
\begin{equation} \label{E^2}
E_{tot}^2 = p^2c^2 + m_0^2c^4 ~.
\end{equation}
It demonstrates how the velocity-dependence of inertia in (\ref{m0}) leads to the standard special relativistic description using the kinematics in Minkowski space-time, where the description acquires the simplest or mathematically economical form, though at the cost of masking of the underlying Machian physics and the preferred cosmological frame.

We have already asserted earlier in \cite{Gog2} that the Machian model of the universe is also capable of reproducing the standard predictions of General Relativity, as it is compatible with the weak equivalence principle, the local Lorentz invariance, and the local position invariance, which are sufficient to describe gravity in terms of the geometry of space-time \cite{Wil}.

This assertion can be argued as follows. When there exists a point-like source of the gravitational potential $\phi (r) = - Gm/r$, the total gravitational potential $\Phi_{tot}$ at a distance $r$ from $m$ can be written as the sum of $\phi$ and the Machian potential $\Phi$:
\begin{equation} \label{c->v}
\Phi_{tot} = \frac{2GM_U}{R} - \frac{Gm}{r}
= c^2 \left[1 - \frac{\phi (r)}{c^2} \right] ~.
\end{equation}
It can be understood as if the parameter $c$, which is generated by the Machian gravitational potential of the whole universe, changes in the neighborhood of a gravitating body of mass $m$. Because of the local $r$-dependence of $\Phi_{tot}$ all kinds of particles will be universally deflected in the vicinity of a gravitating body.

The formula (\ref{E/Phi}) for the mass of a particle can be generalized now to the case of a particle moving in the field of a gravitational potential $\phi (r)$:
\begin{equation} \label{E/Phitot}
m = \frac{E_{tot}}{\Phi_{tot}} ~.
\end{equation}
In this case the calculations similar to (\ref{cal}) lead to a gravitational modification of the Lorentz factor (\ref{gamma}):
\begin{equation} \label{phi}
\gamma_{tot} = \frac{1}{\sqrt{1 + 2\phi (r) /c^2 - v^2/c^2}}~.
\end{equation}

The form of the local modification of the universal speed $c$ and the Lorentz factor $\gamma_{tot}$ by a gravitating body, (\ref{c->v}) and (\ref{phi}), allows us (in a preferred reference frame) to associate with $\phi (r)$ a radial velocity
\begin{equation}
v^r \sim \sqrt{\phi}
\end{equation}
and the related radial acceleration
\begin{equation}
a^r = \frac{dv^r}{dt}~.
\end{equation}

Correspondingly, in an arbitrary reference frame, any local gravitational potential $\phi (r)$ embedded in the global Machian potential $\Phi$ can be associated with 4-dimensional quantities, such as the 4-velocity
\begin{equation}
u^\alpha = e^{-\phi}\xi^\alpha~
\end{equation}
related to the global time-like Killing 4-vector field $\xi^\alpha$, and the 4-acceleration
\begin{equation} \label{a-nu}
a^\nu = u^\alpha \nabla_\alpha u^\nu = - \nabla^\nu \phi~
\end{equation}
which is perpendicular to the causal horizon of the gravitating source.

Now, let us note that the potential (\ref{c->v}) can be obtained as a solution of the Poisson equation:
\begin{equation} \label{Poisson}
\triangle \Phi_{tot} = \triangle \phi = 4\pi G \rho ~,
\end{equation}
where $\rho = dm/dV$ is the mass density. Using the Gauss law this equation can be written in the integral form:
\begin{equation}
E_{tot} = mc^2 = \frac{c^2}{4\pi G} \oint  e^\phi N^\nu \nabla_\nu \phi dS~,
\end{equation}
where $dS$ is the surface element of the causal horizon and $N^\nu$ is the normal to it. Then, using the 4-dimensional formalism, the left hand side of this equation can be identified with the Komar mass and expressed as an integral of a certain combination of the stress energy tensor $T_{\mu\nu}$, while the right hand side can be written in terms of the Killing vector field $\xi^\alpha$ \cite{Wal},
\begin{equation}\label{T-T}
2\int \left( T_{\mu\nu} - \frac 12 g_{\mu\nu}T \right) N^\mu\xi^\nu dV =
\frac{c^4}{8\pi G} \oint \epsilon _{\alpha\beta\mu\nu}\nabla^\mu\xi^\nu dx^\alpha\wedge dx^\beta~,
\end{equation}
where $V$ is the volume inside of the horizon. Finally, using the geometric relation
\begin{equation}
R^\mu_{~\nu} \xi^\nu = - \nabla^\nu \nabla_\nu \xi^\mu
\end{equation}
and the Stokes theorem for the surface integration term in (\ref{T-T}), we obtain the Einstein equation
\begin{equation}
R_{\mu \nu} = \frac{8\pi G}{c^4} \left( T_{\mu\nu} - \frac 12
g_{\mu\nu}T \right) ~.
\end{equation}
Let us underline that this equation is obtained here from the modified Machian energy balance equation, (\ref{E/Phitot}), which singles out a preferred cosmological frame described by $v^i = 0$. The 4-dimensional formalism emerges as a description, which is useful only for isolated systems, when the Machian influence of the universe can be effectively taken into account by means of a local modification of kinematics, i.e. the geometry of space-time.


\section{Thermodynamic Gravity}

We are ready now to describe the Machian universe using the thermodynamic language.

It is known that entropy is a natural parameter to describe a system when some information about it is missing. For example, the thermodynamic description of black holes is related to the lack of information about the interior of the black hole for the outer observers. In our Machian model there are several sources of uncertainties and hence, entropy. First, since we have assumed that all the particles in the universe are "gravitationally entangled", we may expect something similar to the entanglement entropy which is related to the observer's partial lack of knowledge about the quantum state of the system \cite{Entang}. Second, the expansion of the universe and the appearance of the fundamental velocity, (\ref{Phi}), create an event horizon, since $c$ and $H$ define a minimal time of the Machian non-local response of the universe, $\Delta t$. Moreover, the assumption (\ref{A}) is actually equivalent to the fundamental uncertainty relation \cite{Gog3}, which is the source of randomness in quantum mechanics. Indeed, any interaction process of a particle involves at least one another particle from the world ensemble that will contribute an extra term $p_idx^i$ to (\ref{A}) and hence, add at least one extra action quantum $\hbar$.

Using the analogies with the black hole thermodynamics \cite{Haw} applied to the whole Universe let us write the Machian energy of a particle, (\ref{mPhi}), i.e. the energy of the particle's gravitational interaction with the world ensemble, in terms of the temperature $T$ and entropy $S$:
\begin{equation} \label{E=TS}
mc^2 = - m\Phi = TS ~.
\end{equation}
Let us note that additional terms including pressure and chemical potential are absent in (\ref{E=TS}) because our model world ensemble consists of identical particles and there are no natural borders of the world ensemble to create a pressure on. So we use only the simplified thermodynamic relation (\ref{E=TS}), which relates the energy (or mass) of a particle to the variation of temperature and entropy of the world ensemble caused by the particle.

The elementary action of a typical particle, (\ref{A}), can now be written in terms of the thermodynamic variables:
\begin{equation} \label{TS/omega}
A = \hbar = \frac{TS}{2\pi \omega}~,
\end{equation}
where the quantity
\begin{equation}
\omega = \frac{1}{2\pi\Delta t}~,
\end{equation}
is introduced, which can be understood as a fundamental frequency of oscillations of a particle due to its interaction with the world ensemble.

By equating the kinetic energy per degree of freedom with the average kinetic energy of a particle-oscillator from the world ensemble:
\begin{equation}\label{kT}
\frac{kT}{2} = \frac{\omega \hbar}{2} ~,
\end{equation}
where $k$ is the Boltzmann constant, we obtain from (\ref{TS/omega}) an expression for the entropy per typical particle in our Machian universe:
\begin{equation} \label{S=k}
S = 2\pi k~,
\end{equation}
which accounts for an inherent uncertainty characterizing the particle interacting with the world ensemble. Note that the expression (\ref{S=k}) coincides with the formula of entropy used by Verlinde in \cite{Ver}, and it has an extra factor $2\pi$ as compared with the Bekenstein entropy.

Let us consider the energy needed to displace a particle of the world ensemble to the characteristic distance $\Delta x = c \Delta t$:
\begin{equation} \label{DeltaE}
\Delta E = F \Delta x = T \Delta S~,
\end{equation}
where $F = m a$ is a force acting on the particle. Since this is the elementary act of interaction of a single particle with the world ensemble, from (\ref{S=k}) we find:
\begin{equation}
\Delta S = 2\pi k~.
\end{equation}
Then, using (\ref{A}), we obtain from (\ref{DeltaE}) the standard Unruh formula relating the temperature and acceleration of a particle \cite{Unr}:
\begin{equation} \label{Unruh}
a_U = \frac{2\pi k c}{\hbar} T~.
\end{equation}
Note that by describing any acceleration in the universe, (\ref{a-nu}), in terms of the local change of temperature and performing the calculations similar to those in the previous section, we could obtain the thermodynamic form of the Einstein equations \cite{Ein,Pad}.

Next, we can obtain a formula for the Machian force in terms of thermodynamic quantities. Let us write a more general form of the energy balance equations (\ref{mPhi}) and (\ref{Etot}) by including the term $U$ corresponding to local interactions (for example, due to the local gravitational potential $\phi(r)$ which was considered in the previous section):
\begin{equation} \label{dW}
-m d\Phi = c^2 dm - d E_{kin} - dU = -T dS~,
\end{equation}
where $d \Phi$ and $dU = - F_i dx^i$ are, respectively, the work done by the 'universal' gravitational potential $\Phi$ and the local forces $F_i = m a_i$ acting on a particle along the distance $dx^i$ ($i = 1,2,3$). It is clear that, together with (\ref{dE}), which we have shown to guarantee the local Lorentz invariance, the relation (\ref{dW}) leads to the expression:
\begin{equation} \label{TdS}
m d\Phi = F_i dx^i = -T dS~.
\end{equation}
This formula is equivalent to the description of Newtonian gravity as an entropic force introduced in \cite{Ver}:
\begin{equation} \label{F=TS/x}
F_i = -T \frac{dS}{dx^i}~.
\end{equation}

Let us recall that the main problems with the idea of gravity as an entopic force were: (i) the concept of holographic screen where the entropy $S$ is been calculated and (ii) the interpretation of $dx^i$, which was introduced in \cite{Ver} as a distance to this screen. In our approach, the holographic surface is the causal horizon of a particle and $dx^i$ is the distance between the surfaces of the causal horizons of interacting particles, which coincides with the distance between the centers of the corresponding horizon spheres of the particles, i.e. with the distance between the interacting particles themselves (see Fig. 1).


\begin{figure}[ht]
\begin{center}
\includegraphics[width=10cm]{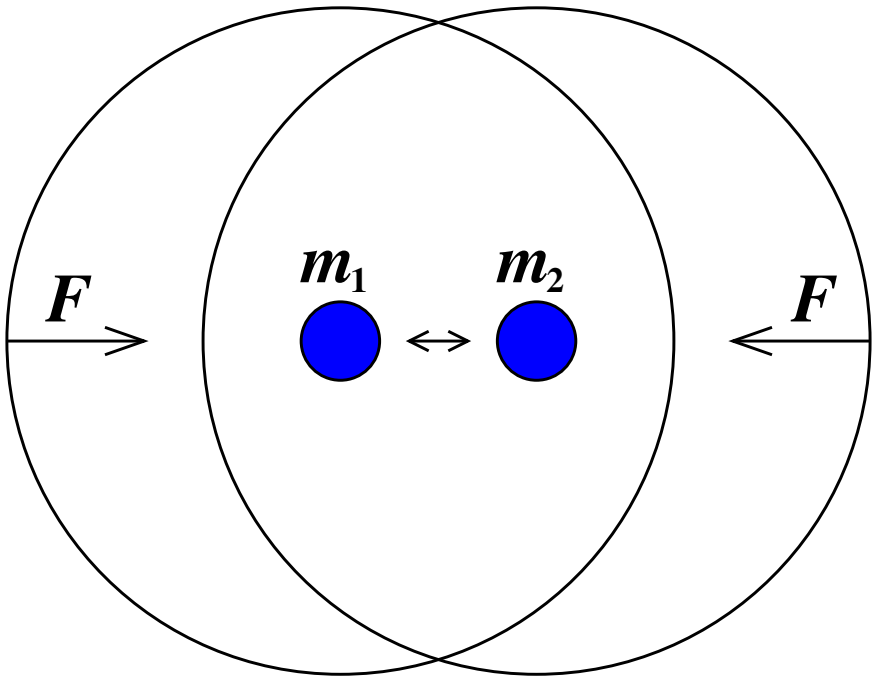}
\caption{Attraction of $m_1$ and $m_2$ via the force $F$ on their horizons}
\end{center}
\end{figure}


Note that some authors have proposed to identify the holographic screen with the Rindler horizon for the accelerated observer \cite{Cul}. However, in this case a derivation of Newton's second law (\ref{F=TS/x}) for circular motion is problematic \cite{DMS}, because the Unruh law (\ref{Unruh}) may have to be modified for rotating observers \cite{Ak-Si}. Usually the discussions of the circular Unruh effect assume that in the Newtonian approximation the mass parameter is a constant. However, in our approach, kinematics and hence, definitions of mass ((\ref{E/Phi}) and (\ref{E/Phitot})) depend on the total energy balance of the physical system with the finite Universe (see Sec. 3), which are different for linear and circular motion. According to (\ref{E=TS}) this difference in energy should result in different effective temperatures for the linearly accelerated and, respectively, rotating observers. The situation is reminiscent of that in general relativity where, due to the problems with definition of the energy of gravitational field, the value of the Schwarzschild mass parameter may depend on the choice of coordinates if the boundary conditions are not fixed properly \cite{Log}.


\section{Temperature of the Universe}

Using the relation between the mass and action of the universe, (\ref{A_U}), we can calculate the entropy of the whole universe:
\begin{equation} \label{SU}
S_U = \frac{N^3}{2} S = \pi k N^3~.
\end{equation}
The formula tells us that even a small uncertainty for a single particle, which is given by (\ref{S=k}), causes an enormous uncertainty for the whole universe.

Equations (\ref{SU}) and and (\ref{S=k}) also lead to the estimation of the total number of bits of information stored at the horizon of the Machian universe consisting of $N$ identical particles:
\begin{equation}
N_{bits} \sim \frac{S_U}{2\pi k} = \frac{N^3}{2}~.
\end{equation}
Note that the appearance of $N^3$ in this formula for the number of bits associated with $N$ identical particles may explain why emerged geometry of the universe is 3-dimensional.

Now, let us write down the energy balance equation, (\ref{E=TS}), for the whole universe:
\begin{equation} \label{TS}
M_U c^2 = \frac {Rc^4}{2G} = T_U S_U ~,
\end{equation}
where $T_U$ is the temperature on the universe's horizon, which can be estimated using (\ref{kT}), (\ref{M}) and (\ref{SU}):
\begin{equation}\label{T_U}
T_U \sim \frac{\omega \hbar }{kN}~ \sim \frac{\hbar H}{2\pi k}
\sim \frac{T}{N} \sim 3\times 10^{-30} ~K~ .
\end{equation}
The value obtained here is of the order of magnitude of the de Sitter temperature \cite{de}. Then the corresponding Unruh acceleration of the horizon:
\begin{equation} \label{a}
a_H = \frac{2\pi c k}{\hbar} T_U = c H \sim 10^{-9}~ m~s^{-2} ~,
\end{equation}
appears to be in good agreement with the observed cosmic acceleration. This result is consistent with the remark in the end of Sect. $2$, that the Machian energy balance condition extrapolated to the whole universe is equivalent to the ($w=-1$) equation of state for the dark energy in standard cosmology.

Finally, by inserting the expression for the temperature at the horizon, (\ref{T_U}), into (\ref{TS}), we arrive at the analogue of the holographic formula for the entropy of the universe
\begin{equation} \label{Su}
S_U = \frac{\pi k c^3}{G \hbar} R^2~,
\end{equation}
according to which $S_U$ is proportional to the quarter of the Hubble horizon area, $4\pi R^2$, similar to the standard holographic formula for the Schwarzschild black holes \cite{Haw}. Let us note the consistency of the scaling behavior of entropy $\sim N^3$ in (\ref{SU}) with the scaling behavior of entropy as an area in (\ref{Su}), since
\begin{equation}
\frac {R^2}{\hbar} \sim N^3~
\end{equation}
as a consequence of (\ref{N}).


\section{Conclusions}

We have shown that the Machian energy balance equations allow us to reproduce relativistic effects within the Machian model of the universe in spite of the fact that it possesses a preferred cosmological reference frame. Simple thermodynamic considerations based on those Machian energy balance equations allow us to define and estimate the entropy of the universe and the Hawking-Unruh temperature at the universe's horizon. We found that this temperature turns out to be consistent with the observed cosmic acceleration and the dark energy density. We also have demonstrated how gravity emerges as an entropic force within the Machian approach and clarified some difficulties in the Verlinde's original consideration.

In the context of this paper it is worthwhile to mention a related work by other authors, namely the recent proposal of a cosmological model where there is no dark energy and the cosmic acceleration is related to a non-zero temperature of the horizon \cite{EFS}, and the attempts to connect the dark energy with quantum uncertainty relations (see \cite{Maz} and the references therein). An interpretation of dark energy as a Machian energy, which is complementary to the "holographic" interpretation obtained in this paper, has been discussed in our previous paper \cite{Gog4}. Besides, our starting assumption of the "gravitationally entangled" Machian world ensemble resembles the idea behind Smolin's recent "real ensemble interpretation of quantum mechanics" \cite{Smo}, which postulates the existence of a world statistical ensemble associated to a quantum state.

\medskip


\noindent {\bf Acknowledgments:} M.G. was supported by by the grant of Shota Rustaveli National Science Foundation $ST~09.798.4-100$. I.K. thanks National Quantum Information Centre of Gda\'nsk (KCIK) for kind hospitality.


\end{document}